\documentclass[conference,10pt]{IEEEtran}
\IEEEoverridecommandlockouts
\usepackage[utf8]{inputenc}
\usepackage{amsmath,epsfig} 
\usepackage{amssymb,amsthm}

\usepackage{url}
\usepackage{epstopdf}
\usepackage{multirow}
\usepackage{setspace}
\usepackage[noadjust]{cite}
\usepackage{xspace}
\usepackage{tabularx}
\usepackage{changepage}
\usepackage{color}
\usepackage{graphicx,bm}
\usepackage{color,cite}
\usepackage{times}
\usepackage{enumerate,type1cm}
\usepackage{amsfonts,relsize}
\usepackage{relsize}
\usepackage{fancybox}
\usepackage[english]{babel}
\usepackage{fancybox}
\usepackage{bbm}
\usepackage{tikz}
\usetikzlibrary{fit,positioning}
\usetikzlibrary{bayesnet}
\usepackage{orcidlink}
\usepackage{caption}

\def\BibTeX{{\rm B\kern-.05em{\sc i\kern-.025em b}\kern-.08em
		T\kern-.1667em\lower.7ex\hbox{E}\kern-.125emX}}

\newtheorem{thm}{\textbf{Theorem}}

\long\def\symbolfootnote[#1]#2{\begingroup%
	\def\thefootnote{\fnsymbol{footnote}}\footnote[#1]{#2}\endgroup}

\newcommand{\beq}{\begin{equation}}
\newcommand{\eeq}{\end{equation}}
\newcommand{\beqa}{\begin{eqnarray}}
\newcommand{\eeqa}{\end{eqnarray}}
\usepackage{float}
\usepackage{colortbl}

\tikzset{
	startstop/.style={
		rectangle, 
		rounded corners,
		minimum width=3cm, 
		minimum height=0.5cm,
		align=center, 
		draw=black, 
	},
	process/.style={
		rectangle, 
		minimum width=3cm, 
		minimum height=0.5cm, 
		align=center, 
		draw=black, 
	},
	decision/.style={
		rectangle, 
		minimum width=3cm, 
		minimum height=0.5cm, align=center, 
		draw=black, 
	},
	arrow/.style={thick,->,>=stealth},
	dec/.style={
		ellipse, 
		align=center, 
		draw=black, 
	},
}


\makeatletter
\def\BState{\State\hskip-\ALG@thistlm}
\makeatother

	{\end{adjustwidth}}

\newcommand{\widesim}[2][1.5]{
  \mathrel{\overset{#2}{\scalebox{#1}[1]{$\sim$}}}
}

\usepackage{array}

\makeatletter
\newcommand{\removelatexerror}{\let\@latex@error\@gobble}
\makeatother

\allowdisplaybreaks

\usepackage{cite}
\usepackage{amsmath,amssymb,amsfonts,amsthm}
\usepackage{algorithmic}
\usepackage{graphicx}
\usepackage{textcomp}
\usepackage{xcolor}
\usepackage{graphicx}
 \usepackage{marginnote}
 \usepackage{fancyhdr}
 \usepackage{epstopdf}
\usepackage{anyfontsize}
\usepackage{makeidx}
\usepackage[utf8]{inputenc}
\usepackage{multirow}
\usepackage{multicol}
\usepackage[none]{hyphenat}
\usepackage{lastpage}
\usepackage[T1]{fontenc}
\usepackage{url} 
\usepackage{epstopdf}
\usepackage{booktabs}
\usepackage{multicol}
\usepackage{fancyhdr}
\usepackage{indentfirst}
\usepackage{color}
\usepackage[section]{placeins}
\usepackage{float}
\usepackage[font=small]{caption}
\usepackage{subcaption}
\usepackage{amsmath, amsthm, amssymb}
\usepackage{bm,bbm}
\usepackage{commath}
\usepackage[ruled,vlined,linesnumbered]{algorithm2e}
\usepackage{stfloats}
\usepackage{afterpage}
\usepackage{etoolbox}
\graphicspath{{Figures/}}
\newcommand{\RNum}[1]{\uppercase\expandafter{\romannumeral #1\relax}}
\def\BibTeX{{\rm B\kern-.05em{\sc i\kern-.025em b}\kern-.08em
    T\kern-.1667em\lower.7ex\hbox{E}\kern-.125emX}}

\usepackage[markup=underlined]{changes}

\definechangesauthor[color=red]{CM}
\definechangesauthor[color=blue]{CRS}
\usepackage{todonotes}
\setcommentmarkup{\todo[color={authorcolor!20},size=\scriptsize]{#3: #1}}

\title{Performance Analysis of Irregular Repetition Slotted Aloha with Multi-Cell Interference 
}

\author{\IEEEauthorblockN{Chirag Ramesh Srivatsa, \IEEEmembership{Graduate Student Member, IEEE,} \orcidlink{0000-0002-3732-4733} and Chandra R. Murthy, \IEEEmembership{Senior Member, IEEE} \orcidlink{0000-0003-4901-9434}} 
\IEEEauthorblockA{{Dept. of CPS and Dept. of ECE,  Indian Institute of Science, Bangalore, India (e-mail: \{chiragramesh, cmurthy\}@iisc.ac.in).}}  
\thanks{This work was financially supported by the Qualcomm Innovation Fellowship and the Young Faculty Research Grant from MeitY, Govt. of India.}
}

\makeatletter
\patchcmd{\@maketitle}
  {\addvspace{0.5\baselineskip}\egroup}
  {\addvspace{-0.9\baselineskip}\egroup}
  {}
  {}
\makeatother

\begin{document}
\maketitle
\begin{abstract}
Irregular repetition slotted aloha (IRSA) is a massive random access protocol in which users transmit several replicas of their packet over a frame to a base station. Existing studies have analyzed IRSA in the single-cell (SC) setup, which does not extend to the more practically relevant multi-cell (MC) setup due to the inter-cell interference. In this work, we analyze MC IRSA, accounting for pilot contamination and multiuser interference. Via numerical simulations, we illustrate that, in practical settings, MC IRSA can have a drastic loss of throughput, up to 70\%, compared to SC IRSA. Further, MC IRSA requires a significantly higher training length (about 4-5x compared to SC IRSA), in order to support the same user density and achieve the same throughput. We also provide insights into the impact of the pilot length, number of antennas, and signal to noise ratio on the performance of MC IRSA.
\end{abstract}


\section{Introduction} \label{sec_intro}
Massive machine-type communications (mMTC) require random access protocols that serve large numbers of users \cite{ref_chen_jsac_2021,ref_mahmood_6G_2020}.
One such protocol is irregular repetition slotted aloha (IRSA),  a successive interference cancellation (SIC) aided protocol in which users transmit multiple packet replicas in different resource blocks (RBs)~\cite{ref_liva_toc_2011}.
Channel estimation in IRSA is accomplished using training or pilot sequences transmitted by the users at the start of their packets. Assigning mutually orthogonal pilots to users avoids pilot contamination, but is prohibitive in mMTC, since the pilot overhead would be proportional to the total number of users~\cite{ref_wu_wc_2020}.
Thus,  \emph{pilot contamination (PC)}, which reduces the accuracy of channel estimation and makes the estimates correlated~\cite{ref_krishnan_twc_2014}, is unavoidable in mMTC, and significantly degrades the throughput of IRSA. 
PC is caused by both  within-cell and out-of-cell users, termed intra-cell PC and inter-cell PC, respectively.
The goal of this paper is to analyze the performance of IRSA, accounting for both intra-cell PC and inter-cell PC.

Initial studies on IRSA with focused on MAC~\cite{ref_liva_toc_2011} and  path loss channels~\cite{ref_khaleghi_pimrc_2017}.
IRSA has been analyzed in a single-cell (SC) setup, accounting for intra-cell PC, estimation errors, path loss, and MIMO fading \cite{ref_srivatsa_spawc_2019,ref_srivatsa_chest_arxiv_2021}.
Multi-user interference from users within the same cell is termed intra-cell interference and from users across cells is termed inter-cell interference.
In the SC setup,  only intra-cell interference affects the decoding of users since users do not face inter-cell interference.
In practice, multiple base stations (BSs) are deployed to cover a large region, and thus inter-cell interference is inevitable~\cite{ref_jose_twc_2011}.
Furthermore,  MC processing (e.g., MC MMSE combining of signals) schemes can achieve better performance compared to SC processing, since it accounts for inter-cell interference~\cite{ref_bjornson_mimo_2017}.


Our main contributions in this paper are as follows: 
\begin{enumerate}
\item We derive the channel estimates in MC IRSA accounting for path loss, MIMO fading,  intra-cell PC, and inter-cell~PC.
\item We analyze the SINR achieved in MC IRSA, accounting for PC,  channel estimation errors, intra-cell interference, and inter-cell interference.
\item We provide insights into the effect of system parameters such as number of antennas, pilot length, and SNR on the throughput performance of MC IRSA.
\end{enumerate}

To the best of our knowledge, no existing work has analyzed the effect of MC interference on IRSA. 
Through numerical simulations, we show that inter-cell PC and inter-cell interference result in up to 70\% loss in throughput compared to the SC setup.
This loss can be overcome by using about $4-5$x longer pilot sequences.
Thus, it is vital to account for the effects of MC interference, in order to obtain realistic insights into the performance of~IRSA.

\textit{Notation:} The symbols $a$, $ \mathbf{a}$, $\mathbf{A}$, $ [\mathbf{A}]_{i,:}$, $ [\mathbf{A}]_{:,j}$, $ \mathbf{0}_N$, $ \mathbf{1}_N,$ and $\mathbf{I}_N $ denote a scalar, a vector, a matrix, the $i$th row of $\mathbf{A}$, the $j$th column of $\mathbf{A}$, all-zero vector of length $N$, all ones vector of length $N$, and an identity matrix of size $N \times N$, respectively. 
$[\mathbf{a}]_{\mathcal{S}}$ and $[\mathbf{A}]_{:,\mathcal{S}}$ denote the elements of $\mathbf{a}$ and the columns of $\mathbf{A}$ indexed by the set $\mathcal{S}$, respectively.
$\text{diag}(\mathbf{a})$ is a diagonal matrix with diagonal entries given by $\mathbf{a}$.
$[N]$ denotes the set $\{1,2,\ldots,N\}$.
$|\cdot|$, $\|\cdot\|$,  and $ [\cdot]^H $ denote the magnitude (or cardinality of a set), $\ell_2 $ norm, and Hermitian operators. 
%
%

\section{System Model} \label{sec_sys_model}
We consider an uplink MC system with $Q$ cells,  where each cell has an $N$-antenna BS located at its center.
We refer to the BS at the center of the $q$th cell as the $q$th BS.
Every cell has $M$ single antenna users arbitrarily deployed within the cell who wish to communicate with their own BS.
The time-frequency resource is divided into $T$ RBs.
These $T$ RBs are common to all the cells, and thus, a total of $QM$ users contend over the $T$ RBs.
Each user randomly accesses a subset of the available RBs according to the IRSA protocol, and transmit packet replicas in the chosen RBs.
Each replica comprises of a header containing pilot symbols for channel estimation,  and a payload containing data and error correction symbols. 

\begin{figure}[t]
	\centering
	\includegraphics[width=0.34\textwidth]{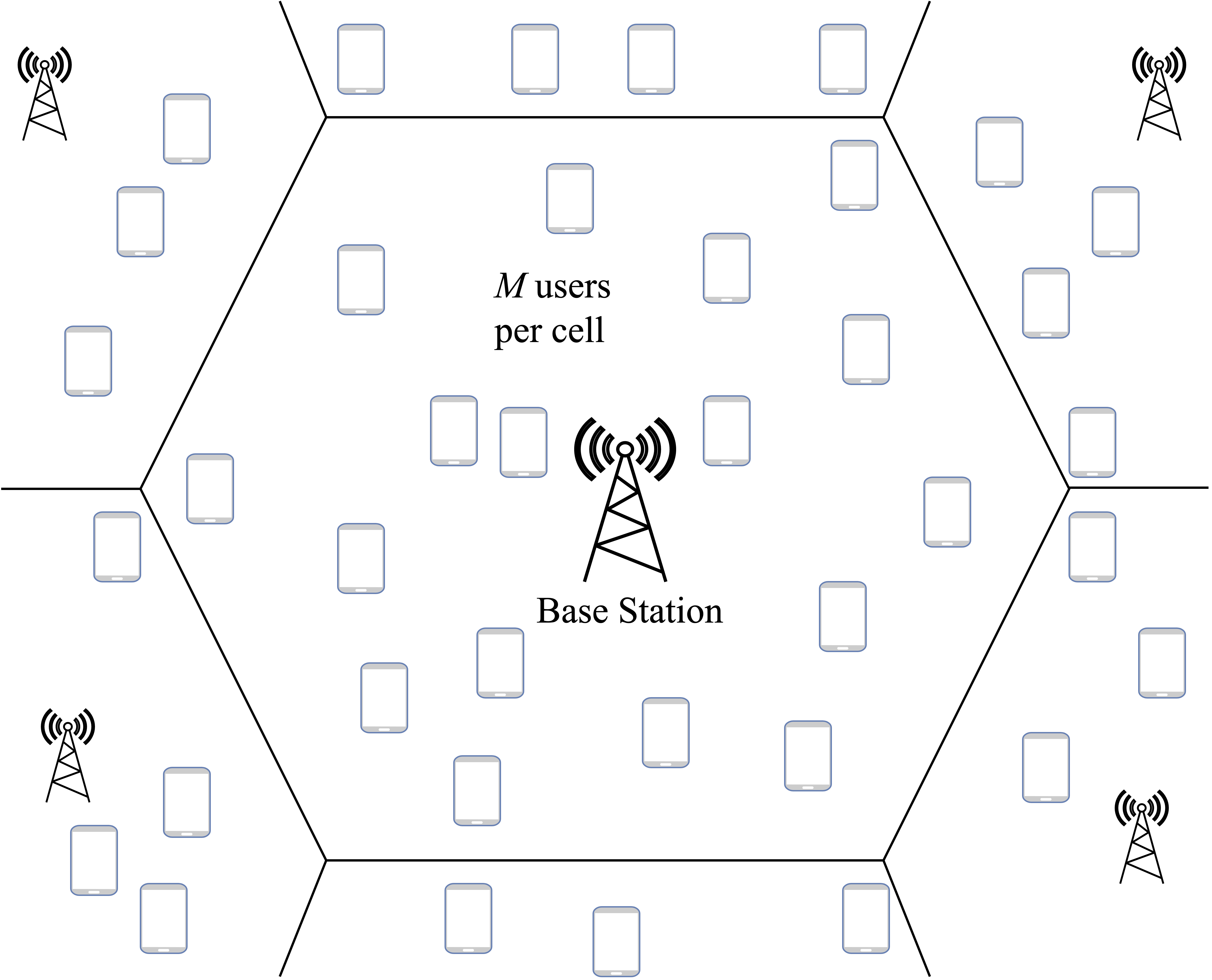}
	\caption{\textcolor{black}{An uplink MC system with $Q$ cells.}}
\label{bs_MC_users_diag_hex}
\end{figure}

The access of the RBs by the users can be represented by an access pattern matrix $\mathbf{G} = [\mathbf{G}_1,\mathbf{G}_2, \ldots, \mathbf{G}_Q ] \in \{0,1\}^{T \times QM} $. 
Here $\mathbf{G}_j \in \{0,1\}^{T \times M}$ represents the access pattern matrix of the users in the $j$th cell,  and 
$g_{tji} = [\mathbf{G}_j ]_{ti}$ is the access coefficient such that $g_{tji} = 1$ if the $i$th user in the $j$th cell transmits in the $t$th RB, and $g_{tji} = 0$ otherwise. 
The $i$th user in the $j$th cell samples its repetition factor $d_{ji}$ from a preset probability distribution. 
It then chooses $d_{ji}$ RBs from the $T$ RBs uniformly at random for transmission. 
The access pattern matrix is known at the BS, which is made possible by using pseudo-random matrices generated from a seed that is available at the BS and the users~\cite{ref_srivatsa_chest_arxiv_2021}.
This can be done in an offline fashion.

The received signal at any BS in the $t$th RB is a superposition of the packets transmitted by the users who choose to transmit in the $t$th RB, across all cells.
In the pilot phase, the $i$th user in the $j$th cell transmits a pilot $ \mathbf{p}_{ji}  \in \mathbb{C}^\tau$ in all the RBs that it has chosen to transmit in, where $\tau$ denotes the length of the pilot sequence.
The received pilot signal at the $q$th BS in the $t$th RB, denoted by $\mathbf{Y}_{tq}^p \in \mathbb{C}^{N \times \tau}$,~is 
\begin{align}
{\mathbf{Y}_{tq}^{p}} &= \textstyle{\sum\nolimits_{j = 1}^Q \sum\nolimits_{i = 1}^M} g_{tji} \mathbf{h}_{tji}^{q} \mathbf{p}_{ji}^{H} + {\mathbf{N}_{tq}^p},
\end{align}
where $\mathbf{N}_{tq}^p \in \mathbb{C}^{N \times \tau}$ is the additive complex white Gaussian noise at the $q$th BS with $[{\mathbf{N}_{tq}^p}]_{nr} \widesim[1.5]{\text{i.i.d.}} \mathcal{CN}(0,N_0)$ $\forall \ n \in [N], \ r \in [\tau]$ and $t \in [T]$, and $N_0$ is the noise variance.
Here,  $\mathbf{h}_{tji}^{q} \in \mathbb{C}^{N}$ is the uplink channel vector between the $i$th user in the $j$th cell and the $q$th BS on the $t$th RB.
The fading is modeled as block-fading, quasi-static and Rayleigh distributed.
The uplink channel is distributed as $\mathbf{h}_{tji}^{q} \widesim[1.5]{\text{i.i.d.}} \mathcal{CN} (\mathbf{0}_N, \beta_{ji}^{q} \sigma_h^2\mathbf{I}_N), \ \forall \ t \in [T], \ i \in [M]$ and $ j \in [Q]$,  where $\sigma_h^2$ is the fading variance, and $\beta_{ji}^{q}$ is the path loss coefficient between the $i$th user in the $j$th cell and the $q$th BS.

In the data phase,  the received data signal at the $q$th BS in the $t$th RB is denoted by $\mathbf{y}_{tq} \in \mathbb{C}^{N}$ and is given by
\begin{align}
{\mathbf{y}_{tq}} &= \textstyle{\sum\nolimits_{j = 1}^Q \sum\nolimits_{i = 1}^M} g_{tji} \mathbf{h}_{tji}^{q} x_{ji}+ {\mathbf{n}_{tq}}, \label{eqn_rec_data}
\end{align}
where $x_{ji}$ is a data symbol with $\mathbb{E}[x_{ji}] = 0$ and $\mathbb{E}[|x_{ji}|^2] = p_{ji}$, i.e., with transmit power $p_{ji}$,  and $\mathbf{n}_{tq} \in \mathbb{C}^N$ is the complex additive white Gaussian noise at the BS, with $[\mathbf{n}_{tq}]_n \widesim[1.5]{\text{i.i.d.}} \mathcal{CN}(0,N_0),$  $\forall \ n \in [N]$ and $t \in [T]$.

\subsubsection{SIC-based Decoding}\label{sec_sic_dec}
%
%
In this work, the decoding of a packet is abstracted into an signal to interference plus noise ratio (SINR) threshold model.
Here, if the SINR of a packet in a given RB in any decoding iteration exceeds a threshold $\gamma_{\text{th}}$, then the packet can be decoded correctly \cite{ref_khaleghi_pimrc_2017,ref_srivatsa_uad_arxiv_2021}.

We now describe the performance evaluation of IRSA via the SINR threshold model. 
In each cell, the BS computes channel estimates and the SINRs of all users in all RBs.
If it finds a user with SINR $\ge \gamma_{\text{th}}$ in some RB, it marks that user's packet as decoded, and performs SIC from all RBs in which the same user has transmitted a replica.
This process of estimation and decoding is carried out iteratively.
Decoding stops when no more users are decoded in two successive iterations.
The throughput is calculated as the number of correctly decoded packets divided by the number of RBs.

\subsubsection{Power Control}
To ensure fairness among users within each cell, we implement a power control policy.
Each user performs path loss inversion with respect to the BS in its own cell~\cite{ref_mirhosseini_tvt_2021}.
That is, the $i$th user in the $j$th cell transmits its symbol $x_{ji}$ at a power $p_{ji} $, i.e.,  $\mathbb{E}[|x_{ji}|^2] = p_{ji} $, according to $p_{ji} = {P}/{\beta_{ji}^{j}}$, where $P$ is a design parameter.
The same power control policy is used in the pilot phase where the transmit power of the $i$th user in the $j$th cell is $p_{ji}^{P} = {P_\tau}/{\beta_{ji}^{j}}$, and $P_\tau \geq P$ is a design parameter, with $ \| \mathbf{p}_{ji}\|^2 = \tau p_{ji}^{P}$.
This ensures a uniform SNR at the BS across all users, with the pilot SNR being $ P_\tau \sigma_h^2/N_0$ and the data SNR being $ P \sigma_h^2/N_0$.
This ensures the power disparity between cell edge users and users located near the BS is reduced, thus ensuring fairness~\cite{ref_mirhosseini_tvt_2021}.

\section{Channel Estimation} \label{sec_ch_est}
Channel estimation is performed based on the received pilot signal in each cell. 
The signals and the channel estimates are indexed by the decoding iteration $k$, since they are recomputed
in every iteration.
We denote the set of users in the $j$th cell who have not yet been decoded up to the $k$th decoding iteration by $\mathcal{S}_{kj}$. For some $m \in \mathcal{S}_{kj}$,  let $\mathcal{S}_{kj}^m \triangleq \mathcal{S}_{kj} \setminus \{m \},$ with $\mathcal{S}_{1j} =  [M]$.
Let the set of all cell indices be denoted by $\mathcal{Q} \triangleq \{1,2,\ldots, Q \}$, and let $\mathcal{Q}^q \triangleq \mathcal{Q} \setminus \{q\}$.
The received pilot signal at the $q$th BS in the $t$th RB in the $k$th decoding iteration is given by
\begin{align}
{\mathbf{Y}_{tq}^{pk}} &= \textstyle{\sum\limits_{i \in \mathcal{S}_{kq} }} g_{tqi} \mathbf{h}_{tqi}^{q} \mathbf{p}_{qi}^{H}  +  \textstyle{\sum\limits_{ j \in \mathcal{Q}^q } \sum\limits_{i \in \mathcal{S}_{1j} }} g_{tji} \mathbf{h}_{tji}^{q} \mathbf{p}_{ji}^{H} + {\mathbf{N}_{tq}^p} , \label{rec_pilot_dec} 
\end{align}
where the first term contains signals from users within the $q$th cell who have not yet been decoded up to the $k$th decoding iteration, i.e., $\forall i \in \mathcal{S}_{kq}$.
The second term contains signals from all users outside the $q$th cell, i.e., from every $ i \in \mathcal{S}_{1j}, \forall j \in \mathcal{Q}^q$.  
We note that there is no coordination among BSs, and thus, 
all the users outside the $q$th cell do not get decoded by the $q$th BS, and they permanently interfere with the decoding of users in other cells, across all the decoding iterations.

Let $\mathcal{G}_{tq} \triangleq \{ i \in  \mathcal{S}_{1q} | g_{tqi} = 1 \}$ denote the set of users within the $q$th cell who have transmitted in the $t$th RB, with $M_{tq} = |\mathcal{G}_{tq}|$.
We denote the set of users in the $q$th cell who have transmitted on the $t$th RB but have not yet been decoded up to the $k$th decoding iteration by $\mathcal{M}_{tq}^{qk} \triangleq \mathcal{G}_{tq}  \cap \mathcal{S}_{kq},$ with ${M}_{tq}^{qk} \triangleq |\mathcal{M}_{tq}^{qk}|$.
Let $\mathbf{H}_{tj}^{q} \triangleq [\mathbf{h}_{tj1}^{q}, \mathbf{h}_{tj2}^{q}, \ldots \mathbf{h}_{tjM}^{q}] $ contain the uplink channels between all the users in the $j$th cell and the $q$th BS, with $\mathbf{H}_{tq}^{qk} \triangleq  [\mathbf{H}_{tq}^{q}]_{:, \mathcal{M}_{tq}^{qk}} $ and $\mathbf{H}_{tj}^{qk} \triangleq  [\mathbf{H}_{tj}^{q}]_{:, \mathcal{G}_{tj}} , \forall j \in \mathcal{Q}^q$.
Let $\mathbf{P}_j \triangleq [\mathbf{p}_{j1}, \mathbf{p}_{j2}, \ldots, \mathbf{p}_{jM}] $ contain the pilots of all users within the $j$th cell, with $\mathbf{P}_{tq}^{qk} \triangleq [\mathbf{P}_{q}]_{:, \mathcal{M}_{tq}^{qk}} $ and $\mathbf{P}_{tj}^{qk} \triangleq [\mathbf{P}_{j}]_{:, \mathcal{G}_{tj}}, \forall j \in \mathcal{Q}^q $.
Let $\mathbf{B}_j^q \triangleq \sigma_h^2 \text{diag} (\beta_{j1}^q, \beta_{j2}^q, \ldots, \beta_{jM}^q ) $ contain the path loss coefficients between the users within the $j$th cell and the $q$th BS, with $\mathbf{B}_{tq}^{qk} \triangleq [\mathbf{B}_{q}^{q}]_{:, \mathcal{M}_{tq}^{qk}} $ and $\mathbf{B}_{tj}^{qk} \triangleq [\mathbf{B}_{j}^{q}]_{:, \mathcal{G}_{tj}}, \forall j \in \mathcal{Q}^q $.
Thus,  the received pilot signal from \eqref{rec_pilot_dec} can be written as 
\begin{align*}
{\mathbf{Y}_{tq}^{pk}} &= \mathbf{H}_{tq}^{qk} \mathbf{P}_{tq}^{qkH} \!\! + \!\! \textstyle{\sum\limits_{ j \in \mathcal{Q}^q }} \mathbf{H}_{tj}^{qk} \mathbf{P}_{tj}^{qkH} \!\! + \! {\mathbf{N}_{tq}^p} = \bar{\mathbf{H}}_{tq}^{qk} \bar{\mathbf{P}}_{tq}^{kH} \! + \! {\mathbf{N}_{tq}^p},
\end{align*}
where $\bar{\mathbf{H}}_{tq}^{qk} \triangleq [\mathbf{H}_{tq}^{qk},  \mathbf{H}_{t1}^{qk}, \ldots, \mathbf{H}_{tq-1}^{qk}, \mathbf{H}_{tq+1}^{qk}, \ldots, \mathbf{H}_{tQ}^{qk}] \in \mathbb{C}^{N \times \bar{M}_{tq}^{qk}}$,  with $\bar{M}_{tq}^{qk} \triangleq {M}_{tq}^{qk} + \sum_{j \in \mathcal{Q}^q} {M}_{tj}$, and $\bar{\mathbf{P}}_{tq}^{k}  \triangleq [\mathbf{P}_{tq}^{qk} ,  \mathbf{P}_{t1}^{qk}, \ldots, \mathbf{P}_{tq-1}^{qk}, \mathbf{P}_{tq+1}^{qk}, \ldots, \mathbf{P}_{tQ}^{qk} ] \in \mathbb{C}^{\tau \times \bar{M}_{tq}^{qk}}$.
We define $\bar{\mathbf{B}}_{tq}^{qk} \triangleq [\mathbf{B}_{tq}^{qk},  \mathbf{B}_{t1}^{qk}, \ldots, \mathbf{B}_{tq-1}^{qk}, \mathbf{B}_{tq+1}^{qk}, \ldots, \mathbf{B}_{tQ}^{qk}] \in \mathbb{C}^{\bar{M}_{tq}^{qk} \times \bar{M}_{tq}^{qk}}$ to derive the channel estimate.
Let $\bar{\mathbf{C}}_{t}^{qk} \! \triangleq \!  \bar{\mathbf{P}}_{tq}^{k} \bar{\mathbf{B}}_{tq}^{qk} ( \bar{\mathbf{P}}_{tq}^{kH} \bar{\mathbf{P}}_{tq}^{k} \bar{\mathbf{B}}_{tq}^{qk}  + N_0 \mathbf{I}_{\bar{M}_{tq}^{qk}})^{-1} \!\! ,$ be split as $ \bar{\mathbf{C}}_{t}^{qk} = [\mathbf{C}_{tq}^{qk}, \mathbf{C}_{t1}^{qk}, \ldots, \mathbf{C}_{tq-1}^{qk}, \mathbf{C}_{tq+1}^{qk}, \ldots, \mathbf{C}_{tQ}^{qk}]  $,
and $\mathbf{c}_{tji}^{qk} \triangleq [{\mathbf{C}}_{tj}^{qk} ]_{:,i}$.

\begin{thm} \label{thm_ch_est}
The minimum mean squared error (MMSE) channel estimate $\hat{\bar{\mathbf{H}}}_{tq}^{qk}$ of $\bar{\mathbf{H}}_{tq}^{qk}$ in the $t$th RB in the $k$th decoding iteration at the $q$th BS can be calculated as
\begin{align}
\hat{\bar{\mathbf{H}}}_{tq}^{qk} 
&= {\mathbf{Y}_{tq}^{pk}}  \bar{\mathbf{P}}_{tq}^{k} \bar{\mathbf{B}}_{tq}^{qk} ( \bar{\mathbf{P}}_{tq}^{kH} \bar{\mathbf{P}}_{tq}^{k} \bar{\mathbf{B}}_{tq}^{qk}  + N_0 \mathbf{I}_{\bar{M}_{tq}^{qk}})^{-1}.
\end{align}
Further,  the estimation error $\tilde{\mathbf{h}}_{tji}^{qk} \triangleq \hat{\mathbf{h}}_{tji}^{qk} - {\mathbf{h}}_{tji}^q$ is distributed as $\tilde{\mathbf{h}}_{tji}^{qk} \sim \mathcal{CN}(\mathbf{0}_N,\delta_{tji}^{qk} \mathbf{I}_N)$, where $\delta_{tji}^{qk}$ is calculated as 
\begin{align*}\delta_{tji}^{qk} = \beta_{ji}^q \sigma_h^2  \left( \!\! \frac{  
\begin{array}{l}
N_0 \|\mathbf{c}_{tji}^{qk}\|^2 
+  \sum\nolimits_{n \in \mathcal{S}_{kj}^i} |\mathbf{p}_{qn}^H \mathbf{c}_{tji}^{qk} |^2  g_{tqn} \beta_{qn}^q \sigma_h^2 \\
+  \sum\nolimits_{l \in \mathcal{Q}^q} \sum\nolimits_{n \in \mathcal{S}_{1j}} |\mathbf{p}_{ln}^H \mathbf{c}_{tji}^{qk} |^2  g_{tln} \beta_{ln}^q \sigma_h^2
\end{array}   }
{ \begin{array}{l}
N_0 \|\mathbf{c}_{tji}^{qk}\|^2 
+  \sum\nolimits_{n \in \mathcal{S}_{kj}} |\mathbf{p}_{qn}^H \mathbf{c}_{tji}^{qk} |^2  g_{tqn} \beta_{qn}^q \sigma_h^2 \\
+  \sum\nolimits_{l \in \mathcal{Q}^q} \sum\nolimits_{n \in \mathcal{S}_{1j}} |\mathbf{p}_{ln}^H \mathbf{c}_{tji}^{qk} |^2  g_{tln} \beta_{ln}^q \sigma_h^2
\end{array}   } \!\! \right) \!\!.
\end{align*}
\end{thm}
\begin{proof}
See Appendix \ref{appendix_ch_est}.
\end{proof}

\noindent \emph{Remark 1:} 
The estimate $\hat{\bar{\mathbf{H}}}_{tq}^{qk}$ can also be calculated as
$\hat{\bar{\mathbf{H}}}_{tq}^{qk} 
= {\mathbf{Y}_{tq}^{pk}}  (\bar{\mathbf{P}}_{tq}^{k} \bar{\mathbf{B}}_{tq}^{qk} \bar{\mathbf{P}}_{tq}^{kH}  + N_0 \mathbf{I}_{\tau})^{-1}  \bar{\mathbf{P}}_{tq}^{k} \bar{\mathbf{B}}_{tq}^{qk},$
(a $\tau \times \tau$ inverse.)
Theorem \ref{thm_ch_est} is applicable for any choice of (possibly non-orthogonal) pilots.
We now discuss the case where pilots are reused by users within and across cells. 

\subsubsection{Pilot Reuse}
Channel estimation is done based on a pilot codebook $\{ \bm \phi_i \}_{i=1}^{\tau}$ of $\tau$ orthogonal pilots~\cite{ref_mirhosseini_tvt_2021}, with each $\bm \phi_i \in \mathbb{C}^\tau$, such that $\bm \phi_i^H \bm \phi_j = 0, \forall i \neq j$, and $\| \bm \phi_i\|^2 = \tau P_\tau$.
Here $P_\tau$ is the pilot power, and the pilot codebook is the same across all cells.
Each user uses a pilot from this codebook, and thus, many users share the same pilot sequence, possibly, both within the cell and out of the cell, 
leading to pilot contamination.
Since $\tau<M$, both intra-cell PC and inter-cell PC occur.

Let $\mathcal{P}_{ji}$ denote the set of users that reuse the pilot of the $i$th user in the $j$th cell.
With this codebook, the channel estimate is distributed as 
$\hat{\mathbf{h}}_{tji}^{qk} \sim \mathcal{CN}(\mathbf{0}_N,\varsigma_{tji}^{qk} \mathbf{I}_N)$, where
$ \varsigma_{tji}^{qk} =  \dfrac{  
\tau P_\tau g_{tji} \beta_{ji}^{q2} \sigma_h^4  }
{ \left( \begin{array}{l}
N_0 +  \sum\nolimits_{n \in \mathcal{S}_{kj} \cap \mathcal{P}_{ji}}  \tau P_\tau g_{tqn} \beta_{qn}^q \sigma_h^2 \\
+  \sum\nolimits_{l \in \mathcal{Q}^q} \sum\nolimits_{n \in \mathcal{S}_{1j} \cap \mathcal{P}_{li}}  \tau P_\tau g_{tln} \beta_{ln}^q \sigma_h^2
\end{array}  \right) } ,$
and the estimation error variance is calculated as
$\delta_{tji}^{qk} = \beta_{ji}^q \sigma_h^2 - \varsigma_{tji}^{qk}.$

\section{Performance Analysis} \label{sec_sinr}
Let $\rho_{tqm}^k $ denote the SINR of the $m$th user in the $q$th cell at the $q$th BS in the $t$th RB in the $k$th decoding iteration.
Similar to \eqref{eqn_rec_data}, the received data signal at the $q$th BS in the $t$th RB in the $k$th decoding iteration is given by
\begin{align}
{\mathbf{y}_{tq}^{k}} &= \textstyle{\sum\limits_{i \in \mathcal{S}_{kq} }} g_{tqi} \mathbf{h}_{tqi}^{q} x_{qi}  +   \textstyle{\sum\limits_{ j \in \mathcal{Q}^q }  \sum\limits_{i \in \mathcal{S}_{1j} }} g_{tji} \mathbf{h}_{tji}^{q} x_{ji} + {\mathbf{n}_{tq}} .
\end{align}
\begin{figure*}[t]
\centering
\begin{align}
\tilde{y}_{tqm}^k = g_{tqm} x_{qm} \mathbf{a}_{tqm}^{kH}  \hat{\mathbf{h}}_{tqm}^{qk} 
- g_{tqm} x_{qm} \mathbf{a}_{tqm}^{kH}  \tilde{\mathbf{h}}_{tqm}^{qk}  
+ \textstyle{\sum\limits_{i \in \mathcal{S}_{kq}^m }} g_{tqi} x_{qi} \mathbf{a}_{tqm}^{kH} \mathbf{h}_{tqi}^{q} 
+ \textstyle{\sum\limits_{ j \in \mathcal{Q}^q }  \sum\limits_{i \in \mathcal{S}_{1j} }} g_{tji}x_{ji} \mathbf{a}_{tqm}^{kH}  \mathbf{h}_{tji}^{q} 
+ \mathbf{a}_{tqm}^{kH} {\mathbf{n}_{tq}}.  \label{eqn_post_comb}  
\end{align}
\vspace{-9mm}
\end{figure*}
We use a combining vector $\mathbf{a}_{tqm}^{k} \in \mathbb{C}^N$ to obtain the post-combined data signal $\tilde{y}_{tqm}^k = \mathbf{a}_{tqm}^{kH} {\mathbf{y}_{tq}^{k}}$ as in \eqref{eqn_post_comb}, with $\tilde{\mathbf{h}}_{tqm}^{qk}$ as defined in Theorem~\ref{thm_ch_est}. 
This combined signal, used to decode the $m$th user in the $q$th cell, is composed of five terms.
The first term $ g_{tqm} x_{qm} \mathbf{a}_{tqm}^{kH}  \hat{\mathbf{h}}_{tqm}^{qk}$ is the useful signal component of the $m$th user; 
the term $g_{tqm} x_{qm} \mathbf{a}_{tqm}^{kH}  \tilde{\mathbf{h}}_{tqm}^{qk} $ arises due to the estimation error $ \tilde{\mathbf{h}}_{tqm}^{k}$; 
the term $\sum\nolimits_{i \in \mathcal{S}_{kq}^m } g_{tqi} x_{qi} \mathbf{a}_{tqm}^{kH} \mathbf{h}_{tqi}^{q}  $ represents the intra-cell interference from the users within the $q$th cell who have transmitted in the $t$th RB and have not yet been decoded up to the $k$th decoding iteration; 
the term $\sum\nolimits_{ j \in \mathcal{Q}^q }  \sum\nolimits_{i \in \mathcal{S}_{1j} } g_{tji}x_{ji} \mathbf{a}_{tqm}^{kH}  \mathbf{h}_{tji}^{q}  $ models the inter-cell interference from users outside the $q$th cell;
and the last term $\mathbf{a}_{tqm}^{kH} {\mathbf{n}_{tq}}$ is the additive noise.
We now present the SINR for all the users.

\begin{thm} \label{thm_sinr}
The signal to interference plus noise ratio (SINR) achieved by the $m$th user within the $q$th cell at the $q$th BS in the $t$th RB and the $k$th decoding iteration can be written as
\begin{align}
\rho_{tqm}^k &\!=\! \dfrac{ {\tt{Gain}}_{tqm}^k }{N_0 + {\tt{InCI}}_{tqm}^k + {\tt{Est}}_{tqm}^k + {\tt{ICI}}_{tqm}^k},  \forall m \in \mathcal{S}_{kq}, \label{eqn_sinr} 
\end{align}
where 
\begin{align*}
{\tt{Gain}}_{tqm}^k &\triangleq p_{qm} g_{tqm}  \textstyle{{ | {\mathbf{a}}_{tqm}^{kH} \hat{\mathbf{h}}_{tqm}^{qk} |^2 }/{\| {\mathbf{a}}_{tqm}^{k} \|^2}} ,  \\
{\tt{InCI}}_{tqm}^k  &\triangleq \textstyle{\sum\nolimits_{i \in \mathcal{S}_{kq}^m}} p_{qi} g_{tqi} { | {\mathbf{a}}_{tqm}^{kH} \hat{\mathbf{h}}_{tqi}^{qk} |^2   }/{  \|{\mathbf{a}}_{tqm}^{k} \|^2}, \\
{\tt{Est}}_{tqm}^k &\triangleq \textstyle{\sum\nolimits_{i \in \mathcal{S}_{kq}}} p_{qi} g_{tqi} \delta_{tqi}^{qk} + \textstyle{\sum\nolimits_{j \in \mathcal{Q}^q} \sum\nolimits_{i \in \mathcal{S}_{1j}}} p_{ji} g_{tji} \delta_{tji}^{qk}  , \\
{\tt{ICI}}_{tqm}^k  &\triangleq \textstyle{\sum\nolimits_{j \in \mathcal{Q}^q} \sum\nolimits_{i \in \mathcal{S}_{1j}}} p_{ji} g_{tji} { | {\mathbf{a}}_{tqm}^{kH} \hat{\mathbf{h}}_{tji}^{qk} |^2   }/{  \|{\mathbf{a}}_{tqm}^{k} \|^2}.
\end{align*}
The channel estimates $\hat{\mathbf{h}}_{tji}^{qk} $ and the error variances  $\delta_{tji}^{qk}$ in the above expressions are obtained from Theorem~\ref{thm_ch_est}. 
\end{thm}
\begin{proof}
See Appendix \ref{appendix_sinr}.
\end{proof}

\noindent \emph{Remark 2:} 
The SINR derived in Theorem \ref{thm_sinr} holds for any choice of the combining vector $\mathbf{a}_{tqm}^{k}$, the pilots, and the power control policy.
The first ${M}_{tq}^{qk}$ columns of the combining matrix $\mathbf{A}_{tq}^{k} \in \mathbb{C}^{N \times \bar{M}_{tq}^{qk}}$ is used at the $q$th BS to decode the ${M}_{tq}^{qk}$ users within the $q$th cell who have not yet been decoded up to the $k$th decoding iteration in the $t$th RB.
The SINR in \eqref{eqn_sinr} is maximized by multi-cell MMSE combining \cite{ref_bjornson_mimo_2017},
under which the optimal combining matrix can be evaluated as
\begin{align}
& \mathbf{A}_{tq}^{k} = \hat{\bar{\mathbf{H}}}_{tq}^{qk} \bar{\mathbf{D}}_{ptq}^{qk} ( (N_0 + {\tt{Est}}_{tm}^k) \mathbf{I}_{\bar{M}_{tq}^{qk}} + \hat{\bar{\mathbf{H}}}_{tq}^{qH} \hat{\bar{\mathbf{H}}}_{tq}^{qk} \bar{\mathbf{D}}_{ptq}^{qk} )^{-1} \!\!\!\!\!\!, \nonumber \\
&= ( (N_0 + {\tt{Est}}_{tm}^k) \mathbf{I}_{N} + \hat{\bar{\mathbf{H}}}_{tq}^{qk} \bar{\mathbf{D}}_{ptq}^{qk} \hat{\bar{\mathbf{H}}}_{tq}^{qH}  )^{-1} \hat{\bar{\mathbf{H}}}_{tq}^{qk} \bar{\mathbf{D}}_{ptq}^{qk}, \nonumber 
\end{align}
where $\mathbf{D}_{pj} \!\! \triangleq \!\! \text{diag} (p_{j1}, p_{j2}, \ldots, p_{jM} ) $ contains the power coefficients of the users within the $j$th cell, $\bar{\mathbf{D}}_{ptq}^{qk} \triangleq [\mathbf{D}_{ptq}^{qk},  \mathbf{D}_{pt1}^{qk}, \ldots, \mathbf{D}_{ptq-1}^{qk}, \mathbf{D}_{ptq+1}^{qk}, \ldots, \mathbf{D}_{ptQ}^{qk}] \in \mathbb{C}^{\bar{M}_{tq}^{qk} \times \bar{M}_{tq}^{qk}}$,  $\mathbf{D}_{ptq}^{qk} \triangleq [\mathbf{D}_{pq}]_{:, \mathcal{M}_{tq}^{qk}} $, and $\mathbf{D}_{ptj}^{qk} \triangleq [\mathbf{D}_{pj}]_{:, \mathcal{G}_{tj}}, \forall j \in \mathcal{Q}^q $.
We note that the above MC processing outperforms the application of SC processing applied to the MC setup~\cite{ref_bjornson_mimo_2017}.

We now present simple and interpretable expressions for the SINR in the massive MIMO (large $N$) regime, and with maximal ratio combining, i.e., $\mathbf{a}_{tqm}^{k} = \hat{\mathbf{h}}_{tqm}^{qk}$~\cite{ref_bjornson_mimo_2017}. 

\begin{thm} \label{thm_sinr_mrc_deteq}
As the number of antennas $N$ gets large, the SINR with maximal ratio combining converges almost surely~to
\begin{align}
\overline{\rho}_{tqm}^k &= \frac{ {\tt Sig}_{tqm}^k }{\epsilon_{tqm}^k ( N_0 + {\tt IntNC}_{tqm}^k ) + {\tt IntC}_{tqm}^k},\label{eqn_sinr_mrc_deteq}
\end{align}
where ${\tt Sig}_{tqm}^k$ is the desired signal, $ {\tt IntNC}_{tqm}^k$ represents the non-coherent interference,  and ${\tt IntC}_{tqm}^k$ represents the coherent interference.
These can be evaluated as
\begin{align*}
\epsilon_{tqm}^k  \! &= \! 
\left(\!\!\!   \begin{array}{l}
N_0 \|\mathbf{c}_{tqm}^{qk}\|^2 
+  \sum\nolimits_{n \in \mathcal{S}_{kj}} |\mathbf{p}_{qn}^H \mathbf{c}_{tqm}^{qk} |^2  g_{tqn} \beta_{qn}^q \sigma_h^2 \\
+  \sum\nolimits_{l \in \mathcal{Q}^q} \sum\nolimits_{n \in \mathcal{S}_{1j}} |\mathbf{p}_{ln}^H \mathbf{c}_{tqm}^{qk} |^2  g_{tln} \beta_{ln}^q \sigma_h^2
\end{array} \!\!\!   \right) \!\!,  \\
{\tt Sig}_{tqm}^k  \! &= \! N p_{qm} g_{tqm} (\epsilon_{tqm}^k)^2 \!\!, \\
{\tt IntNC}_{tqm}^k \!  &= \! \left( \!\!\!  \begin{array}{l}
p_{qm} g_{tqm} \delta_{tqm}^{qk} +  \sum\nolimits_{n \in \mathcal{S}_{kj}}  p_{qn} g_{tqn} \beta_{qn}^q \sigma_h^2 \\
+  \sum\nolimits_{l \in \mathcal{Q}^q} \sum\nolimits_{n \in \mathcal{S}_{1j}} p_{ln} g_{tln} \beta_{ln}^q \sigma_h^2
\end{array} \!\!\!  \right)  \!\!,  \\
{\tt IntC}_{tqm}^k \! &= \! N  \!
\left( \!\! \!\!  \begin{array}{l} 
\sum\nolimits_{n \in \mathcal{S}_{kj}} |\mathbf{p}_{qn}^H \mathbf{c}_{tqm}^{qk} |^2  p_{qn} g_{tqn} \beta_{qn}^{q2} \sigma_h^4 \\
+  \sum\nolimits_{l \in \mathcal{Q}^q} \sum\nolimits_{n \in \mathcal{S}_{1j}} |\mathbf{p}_{ln}^H \mathbf{c}_{tqm}^{qk} |^2 p_{ln} g_{tln} \beta_{ln}^{q2} \sigma_h^4
\end{array} \!\! \! \! \right)  \!\!.
\end{align*}
\end{thm}
\begin{proof}
See Appendix \ref{appendix_sinr_mrc_deteq}.
\end{proof}

\noindent \emph{Remark 3:} 
${\tt IntNC}_{tqm}^k$ represents the non-coherent interference that arises due to estimation errors,  intra-cell interference, and inter-cell interference.
${\tt IntC}_{tqm}^k$ is the coherent interference that arises due to intra-cell PC and inter-cell PC.
The former does not scale with the number of antennas $N$,  whereas the latter scales linearly with $N$. 
Both inter-cell PC and inter-cell interference degrade the performance of the system~\cite{ref_jose_twc_2011}, and thus it is vital to account for both while analyzing the performance of IRSA.

\section{Numerical Results} \label{sec_results}
In this section, the derived SINR analysis is used to evaluate the throughput of MC IRSA via Monte Carlo simulations and provide insights into the impact of various system parameters on the performance of the system.
In each simulation, we generate independent realizations of the user locations, the access pattern matrix, and the channels. 
The throughput in each run is calculated as described in Sec.~\ref{sec_sic_dec},  and the effective system throughput is calculated by averaging over the runs.
We consider a set of $Q=9$ square cells, stacked in a $3 \times 3$ grid,  and report the performance of the center cell~\cite{ref_krishnan_twc_2014}.
Each cell has $M$ users spread uniformly at random across an area of $250 \times 250$~m$^2$,  with the BS at the center~\cite{ref_bjornson_mimo_2017}.\footnote{Due to path loss inversion, the area of the cell does not significantly affect the throughput, but affects the \emph{area spectral efficiency}, which we do not analyze here.}

\begin{figure}[t]
	\centering
	\includegraphics[width=0.5\textwidth]{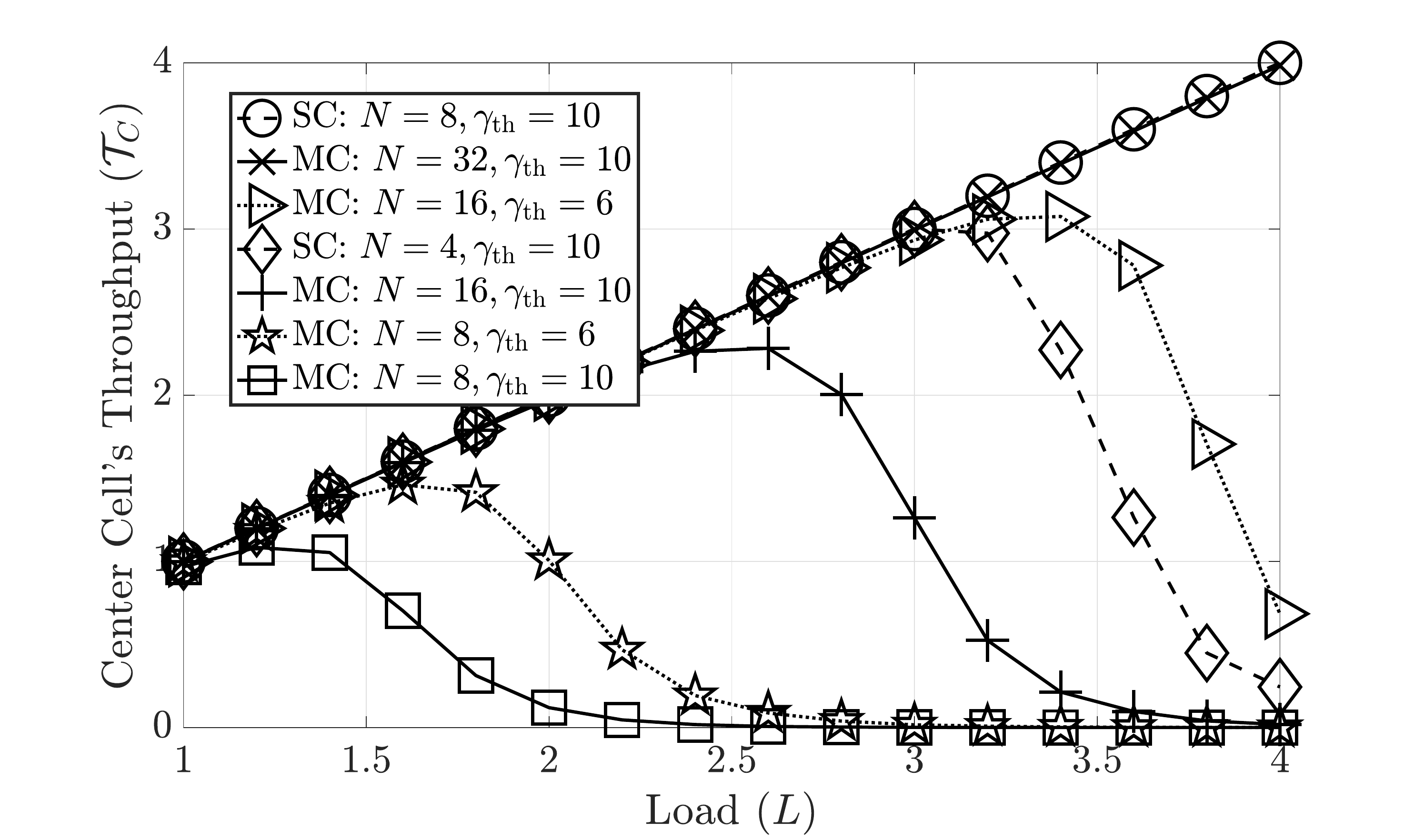}
	\caption{Effect of load $L$ with $\tau=M$.}
\label{fig_mcthpt_vs_L}
\end{figure}
\begin{figure}[t]
	\centering
	\includegraphics[width=0.5\textwidth]{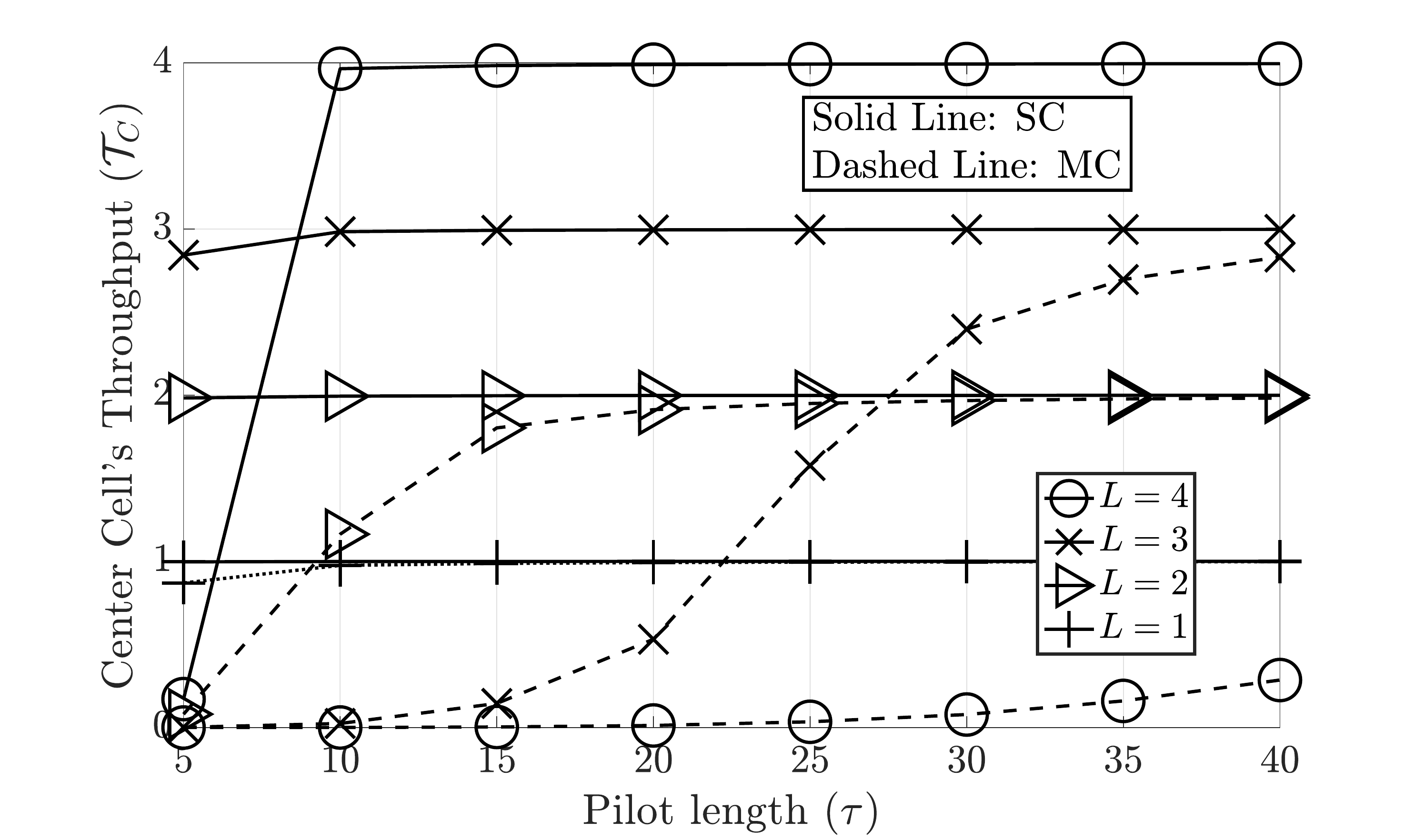}
	\caption{Impact of pilot length $\tau$ with $N=32$.}
\label{fig_mcthpt_vs_tau}
\end{figure}
\begin{figure}[t]
	\centering
	\includegraphics[width=0.5\textwidth]{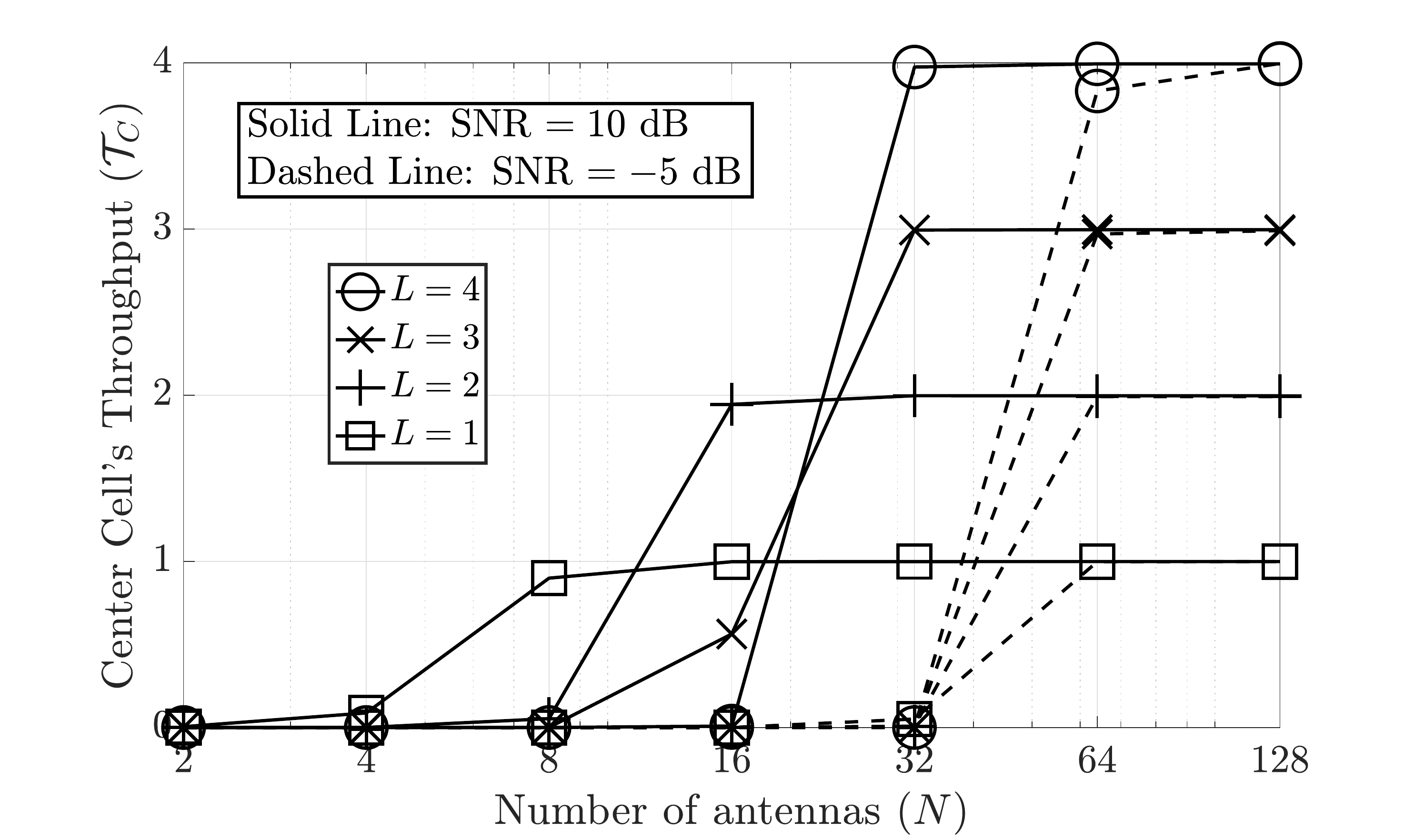}
	\caption{Effect of number of antennas $N$ with $\tau=M$.}
\label{fig_mcthpt_vs_N}
\end{figure}
\begin{figure}[t]
	\centering
	\includegraphics[width=0.5\textwidth]{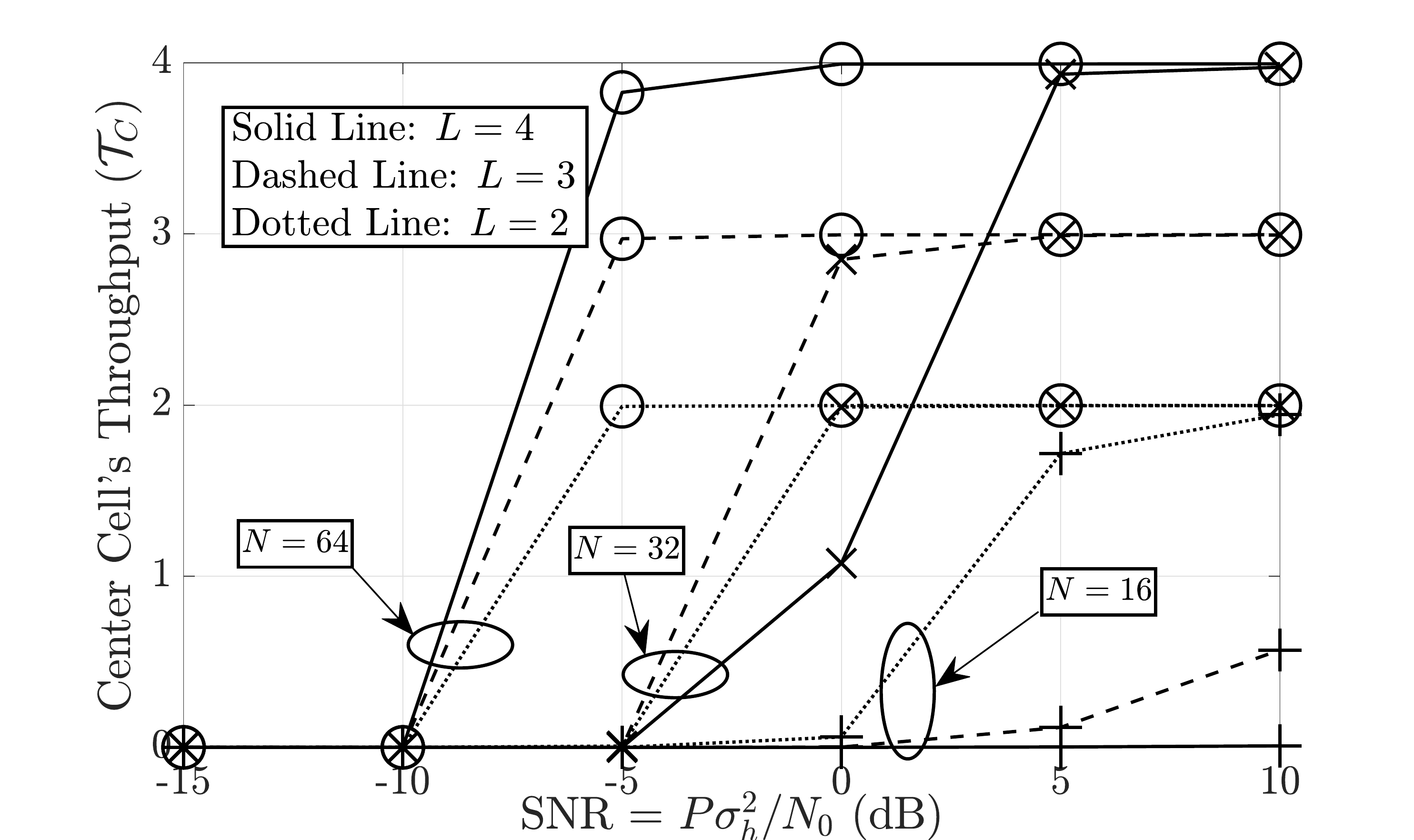}
	\caption{Impact of SNR with $\tau=M$.}
\label{fig_mcthpt_vs_SNR}
\end{figure}

The results in this section are for $T = 50$ RBs, $N_s = 10^3$ Monte Carlo runs, $\sigma_h^2 = 1$, SINR threshold $\gamma_{\text{th}} = 10$.
The number of users contending for the $T$ RBs in each cell is computed based on the load $L$ as $M = \lfloor L T \rceil$. 
The path loss is calculated as $\beta_{ji}^q $ (dB) $= -37.6 \log_{10} (d_{ji}^q/10{\text{m}})$, where $d_{ji}^q$ is the distance of the $i$th user in the $j$th cell from the $q$th BS~\cite{ref_bjornson_mimo_2017}.
The pilot sequences are chosen as the columns of the $\tau \times \tau$ discrete Fourier transform matrix normalized to have column norm $\sqrt{\tau P_\tau}$.
The soliton distribution~\cite{ref_srivatsa_spawc_2019} with $ d_{\max} = 8 $ maximum repetitions is used to generate the repetition factor $d_{ji}$, for the $i$th user in the $j$th cell, whose access vector is formed by uniformly randomly choosing $d_{ji}$ RBs from $T$ RBs without replacement~\cite{ref_liva_toc_2011}.\footnote{The soliton distribution has been shown to achieve 96\% of the throughput that can be achieved with the optimal repetition distribution~\cite{ref_khaleghi_pimrc_2017}.} 
The access pattern matrix is formed by stacking the access vectors of all the users. 
%
The power level is set to $P =P_{\tau} =10$ dBm~\cite{ref_bjornson_mimo_2017} and $N_0$ is chosen such that the data and pilot SNR are $10$ dB, unless otherwise stated.

In Fig. \ref{fig_mcthpt_vs_L}, we show the effect of the load $L$ on the center cell's throughput $\mathcal{T}_C$.
All the curves increase linearly till a peak, which is the desired region of operation, and then drop quickly to zero as the system becomes  interference limited.
All the users' packets are successfully decoded in the linear region of increase, and at high $L$, beyond the peak, the throughputs drop to zero.
For $N=8, \gamma_{\rm th}=10,$ we see a 70\% drop in the peak throughput from $\mathcal{T}_C = 4$ at $L=4$ for SC to $\mathcal{T}_C = 1.2$ at $L=1.2$ for MC. 
This is because users face a high degree of inter-cell interference in the MC setup, unlike the SC setup, especially at high $L$.
In the SC setup, the peak throughput reduces from $\mathcal{T}_C = 4$ for $N=8$ to $\mathcal{T}_C = 3$ at $L=3$ for $N=4$. 
This trend is similar to the MC setup for which the peak throughputs are $\mathcal{T}_C = 4, 2.6, 1.2$ for $N=32,16,8$, respectively.
This is because the system's interference suppression ability with MMSE combining reduces as we decrease $N$~\cite{ref_bjornson_mimo_2017}.
This holds true with $\gamma_{\rm th}=6$ also, which corresponds to a lower SINR threshold, and consequently higher $\mathcal{T}_C$.
To summarize, at high $L$, there is a high degree of inter-cell interference which SC processing does not account for,  resulting in a substantial drop in performance.

Fig. \ref{fig_mcthpt_vs_tau} studies the impact of the pilot length $\tau$.
The performance of SC IRSA at all $L$ is optimal (note that the throughput is upper bounded by $L$) for $\tau >10$.
In MC IRSA, nearly optimal throughputs are achieved for $L=1,2,3$ at $\tau = 10, 30, 40$, respectively.
The throughput for $L=4$ does not improve much with $\tau$.
At high $L$, the impact of inter-cell interference is severe, as expected. 
Increasing $\tau$ implies that each cell has a higher number of orthogonal pilots,  and hence can help in reducing intra-cell PC, but the system is still impacted by inter-cell PC and inter-cell interference.
Thus, we see that MC IRSA requires significantly higher $\tau$ (at least $4-5$x) to overcome inter-cell PC and inter-cell interference to achieve the same performance as that of SC IRSA.

In Fig. \ref{fig_mcthpt_vs_N}, we study the effect of $N$ for $L = 1, 2, 3, 4$, with SNR $=10,-5$ dB.
Nearly optimal throughputs can be achieved with $N=8,16,32,32$ for SNR $=10$ dB, and with $N=64$ for SNR $=-5$ dB.
The system performance improves because of the array gain and higher interference suppression ability at high $N$.
This aids in reducing not only intra-cell interference, but also inter-cell interference.
However, as discussed in \emph{Remark 3}, the SINRs of the users have a coherent interference component that scales with $N$.
Thus, while an increase in $N$ helps reducing intra-cell interference and inter-cell interference, and improves the system performance, it does not reduce intra-cell PC and inter-cell PC.
Similar observations about $N$ can be made where we study the impact of SNR in Fig. \ref{fig_mcthpt_vs_SNR}.
At very low SNR,  the system is noise limited, and increasing $N$ does not help increase the throughput, which is at zero.
For $N=16$, the throughput is always zero and nearly zero for $L=4$ and $L=3$, respectively.
Optimal throughputs are obtained at higher SNRs for $N=32$ and $64$.
Since boosting transmit powers of the users scales both the signal and interference components equally, the SINR does not increase, and therefore the system performance saturates with SNR.
To summarize, increasing $\tau$, $N$, and the SNR can judiciously help reduce the impact of intra-cell PC and inter-cell PC, as well as intra-cell interference and inter-cell interference.


\section{Conclusions} \label{sec_conclusions}
This paper studied the effect of MC interference, namely inter-cell PC and inter-cell interference, on the performance of IRSA.
The users across cells perform path loss inversion with respect to their own BSs and employ a $\tau$-length pilot codebook for channel estimation.
Firstly, the channel estimates were derived, accounting for path loss, MIMO fading,  intra-cell PC, and intra-cell interference.
The corresponding SINR of all the users were derived accounting for channel estimation errors, inter-cell PC, and inter-cell interference.
It was seen that MC IRSA had a significant degradation in performance compared to SC IRSA, even resulting in up to 70\% loss of throughput in certain regimes. 
Recuperating this loss requires at least $4-5$x larger pilot length in MC IRSA to yield the same performance as that of SC IRSA.
Increasing $\tau, N$, and SNR helped improve the performance of MC IRSA.
These results underscore the importance of accounting for multiuser interference in analyzing IRSA in multi-cell settings. 
\textcolor{black}{Future work could include design of optimal pilot sequences to reduce PC and density evolution~\cite{ref_liva_toc_2011} to obtain the asymptotic throughput.}

\vspace{-0.2cm}
\bibliographystyle{IEEEtran}
\bibliography{IEEEabrv,my_refs}

\begin{thebibliography}{10}
\providecommand{\url}[1]{#1}
\csname url@samestyle\endcsname
\providecommand{\newblock}{\relax}
\providecommand{\bibinfo}[2]{#2}
\providecommand{\BIBentrySTDinterwordspacing}{\spaceskip=0pt\relax}
\providecommand{\BIBentryALTinterwordstretchfactor}{4}
\providecommand{\BIBentryALTinterwordspacing}{\spaceskip=\fontdimen2\font plus
\BIBentryALTinterwordstretchfactor\fontdimen3\font minus
  \fontdimen4\font\relax}
\providecommand{\BIBforeignlanguage}[2]{{%
\expandafter\ifx\csname l@#1\endcsname\relax
\typeout{** WARNING: IEEEtran.bst: No hyphenation pattern has been}%
\typeout{** loaded for the language `#1'. Using the pattern for}%
\typeout{** the default language instead.}%
\else
\language=\csname l@#1\endcsname
\fi
#2}}
\providecommand{\BIBdecl}{\relax}
\BIBdecl

\bibitem{ref_chen_jsac_2021}
X.~Chen, D.~W.~K. Ng, W.~Yu, E.~G. Larsson, N.~Al-Dhahir, and R.~Schober,
  ``Massive access for {5G} and beyond,'' \emph{{IEEE} J. Sel. Areas Commun.},
  vol.~39, no.~3, pp. 615--637, 2021.

\bibitem{ref_mahmood_6G_2020}
N.~H. Mahmood, H.~Alves, O.~A. {López}, M.~Shehab, D.~P.~M. Osorio, and
  M.~Latva-Aho, ``Six key features of machine type communication in {6G},'' in
  \emph{2020 6G SUMMIT}, 2020.

\bibitem{ref_liva_toc_2011}
G.~{Liva}, ``Graph-based analysis and optimization of contention resolution
  diversity slotted {ALOHA},'' \emph{{IEEE} Trans. Commun.}, vol.~59, no.~2,
  pp. 477--487, February 2011.

\bibitem{ref_wu_wc_2020}
Y.~{Wu}, X.~{Gao}, S.~{Zhou}, W.~{Yang}, Y.~{Polyanskiy}, and G.~{Caire},
  ``Massive access for future wireless communication systems,'' \emph{{IEEE}
  Wireless Commun.}, vol.~27, no.~4, pp. 148--156, 2020.

\bibitem{ref_krishnan_twc_2014}
N.~Krishnan, R.~D. Yates, and N.~B. Mandayam, ``Uplink linear receivers for
  multi-cell multiuser {MIMO} with pilot contamination: Large system
  analysis,'' \emph{{IEEE} Trans. Wireless Commun.}, vol.~13, pp. 4360--4373,
  2014.

\bibitem{ref_khaleghi_pimrc_2017}
E.~E. {Khaleghi}, C.~{Adjih}, A.~{Alloum}, and P.~{Muhlethaler}, ``Near-far
  effect on coded slotted {ALOHA},'' in \emph{Proc. PIMRC}, Oct 2017.

\bibitem{ref_srivatsa_spawc_2019}
C.~R. {Srivatsa} and C.~R. {Murthy}, ``Throughput analysis of {PDMA/IRSA} under
  practical channel estimation,'' in \emph{Proc. SPAWC}, July 2019.

\bibitem{ref_srivatsa_chest_arxiv_2021}
C.~R. Srivatsa and C.~R. Murthy, ``On the impact of channel estimation on the
  design and analysis of {IRSA} based systems,'' \emph{arXiv:2112.07242}, 2021.

\bibitem{ref_jose_twc_2011}
J.~Jose, A.~Ashikhmin, T.~L. Marzetta, and S.~Vishwanath, ``Pilot contamination
  and precoding in multi-cell {TDD} systems,'' \emph{{IEEE} Trans. Wireless
  Commun.}, vol.~10, no.~8, pp. 2640--2651, 2011.

\bibitem{ref_bjornson_mimo_2017}
E.~{Björnson}, J.~{Hoydis}, and L.~{Sanguinetti}, \emph{Massive {MIMO}
  Networks: Spectral, Energy, and Hardware Efficiency}, 2017.

\bibitem{ref_srivatsa_uad_arxiv_2021}
C.~R. {Srivatsa} and C.~R. {Murthy}, ``User activity detection for irregular
  repetition slotted {ALOHA} based {MMTC},'' \emph{arXiv:2111.06140}, 2021.

\bibitem{ref_mirhosseini_tvt_2021}
F.~Mirhosseini, A.~Tadaion, and S.~M. Razavizadeh, ``Spectral efficiency of
  dense multicell massive {MIMO} networks in spatially correlated channels,''
  \emph{{IEEE} Trans. Veh. Technol.}, vol.~70, pp. 1307--1316, 2021.

\bibitem{ref_couillet_rmt_2011}
R.~Couillet and M.~Debbah, \emph{Random matrix methods for wireless
  communications}.\hskip 1em plus 0.5em minus 0.4em\relax Cambridge University
  Press, 2011.

\end{thebibliography}

\appendices
\renewcommand{\thesectiondis}[2]{\Alph{section}:}
\section{Proof of Theorem \ref{thm_ch_est}} \label{appendix_ch_est}

The minimum mean squared error (MMSE) channel estimate $\hat{\bar{\mathbf{H}}}_{tq}^{qk}$ of the channel matrix $\bar{\mathbf{H}}_{tq}^{qk}$ in the $t$th RB in the $k$th decoding iteration at the $q$th BS can be calculated as
\begin{align}
\hat{\bar{\mathbf{H}}}_{tq}^{qk} 
&= {\mathbf{Y}_{tq}^{pk}}  \bar{\mathbf{P}}_{tq}^{k} \bar{\mathbf{B}}_{tq}^{qk} ( \bar{\mathbf{P}}_{tq}^{kH} \bar{\mathbf{P}}_{tq}^{k} \bar{\mathbf{B}}_{tq}^{qk}  + N_0 \mathbf{I}_{\bar{M}_{tq}^{qk}})^{-1}.
\end{align}

\subsubsection{Channel estimation}
The received signal is first vectorized as
\begin{align}
\overline{\mathbf{y}}_{tq}^k &\triangleq  \text{vec}({\mathbf{Y}}_{tq}^{pk}) = (\bar{\mathbf{P}}_{tq}^{k*} \otimes \mathbf{I}_N) {\mathbf{h}}_{tq}^k +  \overline{\mathbf{n}}_{tq},
\end{align}
where ${\mathbf{h}}_{tq}^k \triangleq \text{vec} (\bar{\mathbf{H}}_{tq}^{qk})$,  $\overline{\mathbf{n}}_{tq} \triangleq \text{vec} (\mathbf{N}_{tq}^p)$, and $\otimes$ is the Kronecker product.
The MMSE estimate is $ \hat{\mathbf{h}}_{tq}^k \triangleq \mathop{{}\mathbb{E}_\mathbf{z}} [ {\mathbf{h}}_{tq}^k ] $, where $\mathbf{z} = \overline{\mathbf{y}}_{tq}^k $. 
The estimation error $\tilde{\mathbf{h}}_{tq} ^{k}$ $\triangleq$ $\hat{\mathbf{h}}_{tq}^{k} - {\mathbf{h}}_{tq}^k $ is uncorrelated with the estimate and with $\mathbf{z} $.
The conditional statistics of a Gaussian random vector $\mathbf{x}$ are 
\begin{align}
\mathop{{}\mathbb{E}_\mathbf{z}} \left[{\mathbf{x}} \right]
&= \mathop{{}\mathbb{E}} \left[{\mathbf{x}} \right] + \mathbf{K}_{\mathbf{x} \mathbf{z}} \mathbf{K}_{\mathbf{z} \mathbf{z}}^{-1} \left( \mathbf{z} - \mathop{{}\mathbb{E}} \left[{\mathbf{z}} \right] \right), \label{eqn_gauss_cond1} \\
\mathbf{K}_{\mathbf{xx} | \mathbf{z}}
&= \mathbf{K}_{\mathbf{x} \mathbf{x}} - \mathbf{K}_{\mathbf{x} \mathbf{z}} \mathbf{K}_{\mathbf{z} \mathbf{z}}^{-1} \mathbf{K}_{\mathbf{z} \mathbf{x}}. \label{eqn_gauss_cond2}
\end{align}
Here, $\mathbf{K}_{\mathbf{x} \mathbf{x}},$ $\mathbf{K}_{\mathbf{xx} | \mathbf{z}},$ and $ \mathbf{K}_{\mathbf{x} \mathbf{z}}$ are the unconditional covariance of $\mathbf{x}$, the conditional covariance of $\mathbf{x}$ conditioned on $\mathbf{z}$, and the cross-covariance of $\mathbf{x} \ \& \ \mathbf{z}$ respectively.
From \eqref{eqn_gauss_cond1}, the MMSE estimate $ \hat{\mathbf{h}}_{tq}^{k} $ of the channel can be evaluated as
\begin{align*}
& \hat{\mathbf{h}}_{tq}^{k}  =  \mathop{{}\mathbb{E}} {[ {\mathbf{h}}_{tq}^k]} +  \mathop{{}\mathbb{E}} {[ {\mathbf{h}}_{tq}^k{\overline{\mathbf{y}}_{tq}^{kH} } ]}  \mathop{{}\mathbb{E}} [ \overline{\mathbf{y}}_{tq}^{k}  {\overline{\mathbf{y}}_{tq}^{kH} } ] ^{-1} ( \overline{\mathbf{y}}_{tq}^{k}  -  \mathop{{}\mathbb{E}} {[\overline{\mathbf{y}}_{tq}^{k} ]} ).
\end{align*}
The terms in the above expression can be calculated as
\begin{align*}
& \mathop{{}\mathbb{E}} {[ {\mathbf{h}}_{tq}^k{\overline{\mathbf{y}}_{tq}^{kH} } ]} = \bar{\mathbf{B}}_{tq}^{qk} \bar{\mathbf{P}}_{tq}^{kT} \otimes \mathbf{I}_N, \\
& \mathop{{}\mathbb{E}} [ \overline{\mathbf{y}}_{tq}^{k}  {\overline{\mathbf{y}}_{tq}^{kH} } ] =  (\bar{\mathbf{P}}_{tq}^{k*} \bar{\mathbf{B}}_{tq}^{qk} \bar{\mathbf{P}}_{tq}^{kT} + N_0 \mathbf{I}_\tau) \otimes \mathbf{I}_N, \\
& \qquad \hat{\mathbf{h}}_{tq}^{k}  = (\bar{\mathbf{B}}_{tq}^{qk} \bar{\mathbf{P}}_{tq}^{kT}( \bar{\mathbf{P}}_{tq}^{k*}  \bar{\mathbf{B}}_{tq}^{qk} \bar{\mathbf{P}}_{tq}^{kT} + N_0 \mathbf{I}_\tau)^{-1} \otimes \mathbf{I}_N) \overline{\mathbf{y}}_{tq}^{k} ,
\end{align*}
and thus, the MMSE estimate $\hat{\bar{\mathbf{H}}}_{tq}^{qk}$ of $\bar{\mathbf{H}}_{tq}^{qk}$ is
\begin{align}
\hat{\bar{\mathbf{H}}}_{tq}^{qk} &=  {\mathbf{Y}_{tq}^{pk}} ( \bar{\mathbf{P}}_{tq}^{k} \bar{\mathbf{B}}_{tq}^{qk} \bar{\mathbf{P}}_{tq}^{kH} + N_0 \mathbf{I}_\tau)^{-1}  \bar{\mathbf{P}}_{tq}^{k} \bar{\mathbf{B}}_{tq}^{qk} \\
&\stackrel{(a)}{=}   {\mathbf{Y}_{tq}^{pk}} \bar{\mathbf{P}}_{tq}^{k} \bar{\mathbf{B}}_{tq}^{qk} ( \bar{\mathbf{P}}_{tq}^{kH}\bar{\mathbf{P}}_{tq}^{k} \bar{\mathbf{B}}_{tq}^{qk}  + N_0 \mathbf{I}_{\bar{M}_{tq}^{qk}})^{-1},
\end{align}
where $(a)$ follows from $(\mathbf{AB} + \mathbf{I})^{-1}\mathbf{A}$ $=$ $ \mathbf{A}(\mathbf{BA} + \mathbf{I})^{-1}$.

\subsubsection{Error variance}
The conditional covariance of ${\mathbf{h}}_{tji}^q$ is calculated conditioned on the knowledge of $\mathbf{z} = \hat{\mathbf{h}}_{tji}^{qk}$. 
Let $\bar{\mathbf{C}}_{t}^{qk} \! \triangleq \!  \bar{\mathbf{P}}_{tq}^{k} \bar{\mathbf{B}}_{tq}^{qk} ( \bar{\mathbf{P}}_{tq}^{kH} \bar{\mathbf{P}}_{tq}^{k} \bar{\mathbf{B}}_{tq}^{qk}  + N_0 \mathbf{I}_{\bar{M}_{tq}^{qk}})^{-1} \!\!$ be split as $ \bar{\mathbf{C}}_{t}^{qk} = [\mathbf{C}_{tq}^{qk}, \mathbf{C}_{t1}^{qk}, \ldots, \mathbf{C}_{tq-1}^{qk}, \mathbf{C}_{tq+1}^{qk}, \ldots, \mathbf{C}_{tQ}^{qk}]  $,
and $\mathbf{c}_{tji}^{qk} \triangleq [{\mathbf{C}}_{tj}^{qk} ]_{:,i}$.
Thus, we can evaluate
\begin{align*}
\mathbf{K}_{{\mathbf{h}}_{tji}^q {\mathbf{h}}_{tji}^q} &= \mathbb{E} [{\mathbf{h}}_{tji}^q  {\mathbf{h}}_{tji}^{qH}] = \beta_{ji}^q \sigma_h^2 \mathbf{I}_N, \\
\mathbf{K}_{{\mathbf{h}}_{tji}^q \mathbf{z}} &= \mathbb{E} [{\mathbf{h}}_{tji}^q \hat{\mathbf{h}}_{tji}^{qkH}] =  \mathbf{p}_{ji}^H \mathbf{c}_{tji}^{qk}  g_{tji} \beta_{ji}^q \sigma_h^2 \mathbf{I}_N, \\ 
\mathbf{K}_{{\mathbf{z}}\mathbf{z}}  \! = &  \! \left(  \!\!  \begin{array}{l}
N_0 \|\mathbf{c}_{tji}^{qk}\|^2 
+  \sum\nolimits_{n \in \mathcal{S}_{kj}} |\mathbf{p}_{qn}^H \mathbf{c}_{tji}^{qk} |^2  g_{tqn} \beta_{qn}^q \sigma_h^2 \\
+  \sum\nolimits_{l \in \mathcal{Q}^q} \sum\nolimits_{n \in \mathcal{S}_{1j}} |\mathbf{p}_{ln}^H \mathbf{c}_{tji}^{qk} |^2  g_{tln} \beta_{ln}^q \sigma_h^2
\end{array}  \!\!  \right) \! \mathbf{I}_N.
\end{align*}
Thus, the conditional covariance is
\begin{align*}
& \mathbf{K}_{{\mathbf{h}}_{tji}^{q}{\mathbf{h}}_{tji}^{q}| \mathbf{z}}
= \mathbf{K}_{{\mathbf{h}}_{tji}^{q} {\mathbf{h}}_{tji}^{q}} - \mathbf{K}_{{\mathbf{h}}_{tji}^{q}\mathbf{z}} \mathbf{K}_{\mathbf{z} \mathbf{z}}^{-1} \mathbf{K}_{\mathbf{z} {\mathbf{h}}_{tji}^{q}}  \triangleq \delta_{tji}^{qk} \mathbf{I}_N,
\end{align*}
where $\delta_{tji}^{qk}$ is calculated as 
\begin{align*}
\delta_{tji}^{qk} = \beta_{ji}^q \sigma_h^2  \left( \!\! \frac{  
\begin{array}{l}
N_0 \|\mathbf{c}_{tji}^{qk}\|^2 
+  \sum\nolimits_{n \in \mathcal{S}_{kj}^i} |\mathbf{p}_{qn}^H \mathbf{c}_{tji}^{qk} |^2  g_{tqn} \beta_{qn}^q \sigma_h^2 \\
+  \sum\nolimits_{l \in \mathcal{Q}^q} \sum\nolimits_{n \in \mathcal{S}_{1j}} |\mathbf{p}_{ln}^H \mathbf{c}_{tji}^{qk} |^2  g_{tln} \beta_{ln}^q \sigma_h^2
\end{array}   }
{ \begin{array}{l}
N_0 \|\mathbf{c}_{tji}^{qk}\|^2 
+  \sum\nolimits_{n \in \mathcal{S}_{kj}} |\mathbf{p}_{qn}^H \mathbf{c}_{tji}^{qk} |^2  g_{tqn} \beta_{qn}^q \sigma_h^2 \\
+  \sum\nolimits_{l \in \mathcal{Q}^q} \sum\nolimits_{n \in \mathcal{S}_{1j}} |\mathbf{p}_{ln}^H \mathbf{c}_{tji}^{qk} |^2  g_{tln} \beta_{ln}^q \sigma_h^2
\end{array}   } \!\! \right) \!\!.
\end{align*}
The conditional autocorrelation follows as
\begin{align*}
\mathbb{E}_{\mathbf{z}}  [{\mathbf{h}}_{tji}^{q} {\mathbf{h}}_{tji}^{qH}] &= \mathbf{K}_{{\mathbf{h}}_{tji}^{q} {\mathbf{h}}_{tji}^{q}| \mathbf{z}} +  \mathbb{E}_{\mathbf{z}}  [{\mathbf{h}}_{tji}^{q}] \mathbb{E}_{\mathbf{z}}  [{\mathbf{h}}_{tji}^{q} ]^H \\
&= \delta_{tji}^{qk} \mathbf{I}_N + \hat{\mathbf{h}}_{tji}^{qk} \hat{\mathbf{h}}_{tji}^{qkH}.
\end{align*} 
The unconditional and conditional means of the estimation error are $\mathbb{E} [\tilde{\mathbf{h}}_{tji}^{qk}] = \mathbb{E} [\hat{\mathbf{h}}_{tji}^{qk}- {\mathbf{h}}_{tji}^{q}] = 0$ and $\mathbb{E}_{\mathbf{z}} [\tilde{\mathbf{h}}_{tji}^{qk}] = \mathbb{E}_{\mathbf{z}} [\hat{\mathbf{h}}_{tji}^{qk} - {\mathbf{h}}_{tji}^{q}] = \hat{\mathbf{h}}_{tji}^{qk} - \hat{\mathbf{h}}_{tji}^{qk} = 0.$
The conditional autocovariance of the error therefore simplifies as
\begin{align*}
& \mathbf{K}_{\tilde{\mathbf{h}}_{tji}^{qk} \tilde{\mathbf{h}}_{tji}^{qk}| \mathbf{z}} = \mathbb{E}_{\mathbf{z}} [\tilde{\mathbf{h}}_{tji}^{qk} \tilde{\mathbf{h}}_{tji}^{qkH}] \\
& \ \ \ =  \mathbb{E}_{\mathbf{z}} [{\mathbf{h}}_{tji}^{q} {\mathbf{h}}_{tji}^{qH}] - \hat{\mathbf{h}}_{tji}^{qk} \hat{\mathbf{h}}_{tji}^{qkH} = \delta_{tji}^{qk} \mathbf{I}_N,
\end{align*}
and thus,  $\delta_{tji}^{qk}$ is also the variance of the estimation error.

\vspace{-0.2cm}
\section{Proof of Theorem \ref{thm_sinr}}
\label{appendix_sinr}


In order to calculate the SINR, we first evaluate the power of the received signal, which is calculated conditioned on the knowledge of the channel estimates $\mathbf{z} \triangleq  \text{vec} (\hat{\bar{\mathbf{H}}}_{tq}^{qk})$ as $\mathbb{E}_{\mathbf{z}} [|\tilde{y}_{tqm}^k|^2] = \mathbb{E}_{\mathbf{z}} [| \sum_{i=1}^5 T_i|^2]$. 
Since noise is uncorrelated with data, $\mathbb{E}_{\mathbf{z}} [T_1 T_5^H]$ $= \mathbb{E}_{\mathbf{z}} [T_2 T_5^H]$ $= \mathbb{E}_{\mathbf{z}} [T_3 T_5^H]  = \mathbb{E}_{\mathbf{z}} [T_4 T_5^H] = {0} $. 
Since MMSE estimates are uncorrelated with their errors \cite{ref_bjornson_mimo_2017}, $\mathbb{E}_{\mathbf{z}} [T_1 T_2^H] = {0}$.
Finding the other components requires $\mathbb{E}_{\mathbf{z}} [x_{ji} x_{jl}]$ for $i \neq l$ which can be found as $\mathbb{E}_{\mathbf{z}} [x_{ji} x_{jl}] = \mathbb{E}_{\mathbf{z}} [x_{ji}] \mathbb{E}_{\mathbf{z}} [x_{jl}] = 0$. 
Thus, all the five terms are uncorrelated and the power in the received signal is just a sum of the powers of the individual components $\mathbb{E}_{\mathbf{z}} [|\tilde{y}_{tqm}^k|^2] =  \sum_{i=1}^5 \mathbb{E}_{\mathbf{z}} [|T_i|^2]$.
We now compute the powers of each of the components.  The useful signal power~is 
\begin{align*}
\mathbb{E}_{\mathbf{z}} [|T_1|^2] &=  \mathbb{E}_{\mathbf{z}} [|\mathbf{a}_{tqm}^{kH}  \hat{\mathbf{h}}_{tqm}^{qk} g_{tqm} x_{qm}|^2] \\ 
&= p_{qm} g_{tqm}^2 |\mathbf{a}_{tqm}^{kH}  \hat{\mathbf{h}}_{tqm}^{qk}|^2.
\end{align*}
The desired gain is written as
\begin{align} \label{eqn_sig_pow}
{\tt{Gain}}_{tqm}^k &\triangleq \frac{ \mathbb{E}_{\mathbf{z}} [|T_1|^2]  }{ \| \mathbf{a}_{tqm}^{k}  \|^2} = p_{qm} g_{tqm}  \frac{ |\mathbf{a}_{tqm}^{kH}  \hat{\mathbf{h}}_{tqm}^{qk}|^2 }{\| \mathbf{a}_{tqm}^{k}  \|^2}.
\end{align}
The power of the estimation error is expressed as
\begin{align*}
\mathbb{E}_{\mathbf{z}} [|T_2|^2]  &= \mathbb{E}_{\mathbf{z}} [|\mathbf{a}_{tqm}^{kH}  \tilde{\mathbf{h}}_{tqm}^{qk} g_{tqm} x_{qm}|^2] \\
& = p_{qm} g_{tqm}^2 \delta_{tqm}^{qk} \|\mathbf{a}_{tqm}^{k}  \|^2.  
\end{align*}
Next, the power of the intra-cell interference term $T_3$  is
\begin{align*}
&\mathbb{E}_{\mathbf{z}} [|T_3|^2]  = \mathbb{E}_{\mathbf{z}} [ | \mathbf{a}_{tqm}^{kH}  \textstyle{\sum\nolimits_{i \in \mathcal{S}_{kq}^m}}  g_{tqi} {\mathbf{h}}_{tqi}^{q} x_{qi} |^2] \nonumber \\
& \ \ \ \ =  \textstyle{\sum\nolimits_{i \in \mathcal{S}_{kq}^m }} p_{qi} g_{tqi}^2 \mathbf{a}_{tqm}^{kH}  \mathbb{E}_{\mathbf{z}}[ {\mathbf{h}}_{tqi}^{q} {\mathbf{h}}_{tqi}^{qH} ] \mathbf{a}_{tqm}^{k} \nonumber \\
& \ \ \ \ = \textstyle{\sum\nolimits_{i \in \mathcal{S}_{kq}^m }}  p_{qi} g_{tqi}^2 \mathbf{a}_{tqm}^{kH}  (\delta_{tqi}^{qk} \mathbf{I}_N + \hat{\mathbf{h}}_{tqi}^{qk} \hat{\mathbf{h}}_{tqi}^{qkH}) \mathbf{a}_{tqm}^{k} \nonumber \\
& \ \ \ \ =  \textstyle{\sum\nolimits_{i \in \mathcal{S}_{kq}^m }}  p_{qi} g_{tqi}^2 ( \|\mathbf{a}_{tqm}^{k}  \|^2   \delta_{tqi}^{qk} + | \mathbf{a}_{tqm}^{kH}  \hat{\mathbf{h}}_{tqi}^{qk}|^2 ).
\end{align*} 
Then, the power of the inter-cell interference term $T_4$  is
\begin{align*}
&\mathbb{E}_{\mathbf{z}} [|T_4|^2]  = \mathbb{E}_{\mathbf{z}} [ | \mathbf{a}_{tqm}^{kH}  \textstyle{\sum\nolimits_{ j \in \mathcal{Q}^q }  \sum\nolimits_{i \in \mathcal{S}_{1j} }}  g_{tji} {\mathbf{h}}_{tji}^{q} x_{ji} |^2] \nonumber \\
& \ \ \ \ = \textstyle{\sum\nolimits_{ j \in \mathcal{Q}^q }  \sum\nolimits_{i \in \mathcal{S}_{1j} }} p_{ji} g_{tji}^2 \mathbf{a}_{tqm}^{kH}  \mathbb{E}_{\mathbf{z}}[ {\mathbf{h}}_{tji}^{q} {\mathbf{h}}_{tji}^{qH} ] \mathbf{a}_{tqm}^{k} \nonumber \\
& \ \ \ \ = \textstyle{\sum\nolimits_{ j \in \mathcal{Q}^q }  \sum\nolimits_{i \in \mathcal{S}_{1j} }} p_{ji} g_{tji}^2 \mathbf{a}_{tqm}^{kH}  (\delta_{tji}^{qk} \mathbf{I}_N + \hat{\mathbf{h}}_{tji}^{qk}  \hat{\mathbf{h}}_{tji}^{qkH}) \mathbf{a}_{tqm}^{k} \nonumber \\
& \ \ \ \ =  \textstyle{\sum\nolimits_{ j \in \mathcal{Q}^q }  \sum\nolimits_{i \in \mathcal{S}_{1j} }} p_{ji} g_{tji}^2 ( \|\mathbf{a}_{tqm}^{k}  \|^2   \delta_{tji}^{qk}  + | \mathbf{a}_{tqm}^{kH}  \hat{\mathbf{h}}_{tji}^{qk} |^2 ).
\end{align*}

Let $\nu =  \mathbb{E}_{\mathbf{z}} [|T_2|^2] + \mathbb{E}_{\mathbf{z}} [|T_3|^2] + \mathbb{E}_{\mathbf{z}} [|T_4|^2]$ represent the joint contribution of estimation errors and multi-user interference components of the other users (both within the $q$th cell and outside the $q$th cell). 
Since $g_{tji} $ is binary, its powers are dropped.
We now split $\nu/  \|\mathbf{a}_{tqm}^{k}  \|^2$ into the sum of the estimation error component ${\tt{Est}}_{tqm}^k$, intra-cell interference ${\tt{InCI}}_{tqm}^k$ and inter-cell interference ${\tt{ICI}}_{tqm}^k $ as follows
\begin{align*}
{\tt{Est}}_{tqm}^k &= \textstyle{\sum\nolimits_{i \in \mathcal{S}_{kq}}} p_{qi} g_{tqi} \delta_{tqi}^{qk} + \textstyle{\sum\nolimits_{j \in \mathcal{Q}^q} \sum\nolimits_{i \in \mathcal{S}_{1j}}} p_{ji} g_{tji} \delta_{tji}^{qk}  , \\
{\tt{InCI}}_{tqm}^k  &= \textstyle{\sum\nolimits_{i \in \mathcal{S}_{kq}^m}} p_{qi} g_{tqi} { | {\mathbf{a}}_{tqm}^{kH} \hat{\mathbf{h}}_{tqi}^{qk} |^2   }/{  \|{\mathbf{a}}_{tqm}^{k} \|^2}, \\
{\tt{ICI}}_{tqm}^k  &= \textstyle{\sum\nolimits_{j \in \mathcal{Q}^q} \sum\nolimits_{i \in \mathcal{S}_{1j}}} p_{ji} g_{tji} { | {\mathbf{a}}_{tqm}^{kH} \hat{\mathbf{h}}_{tji}^{qk} |^2   }/{  \|{\mathbf{a}}_{tqm}^{k} \|^2}.
\end{align*}

The noise power is calculated as
\begin{align} \label{eqn_noise_pow}
\mathbb{E}_{\mathbf{z}} [|T_5|^2]  &= \mathbb{E}_{\mathbf{z}} [|\mathbf{a}_{tqm}^{k}  \mathbf{n}_{tq}|^2] =  N_0 \|\mathbf{a}_{tqm}^{k}  \|^2.
\end{align}
A meaningful SINR expression can be written out by dividing the useful gain from \eqref{eqn_sig_pow} by the sum of the interference and the noise powers (from $\nu$ and \eqref{eqn_noise_pow}) \cite{ref_bjornson_mimo_2017}.
Note that the interference component is comprised of the estimation error term and the signal powers of other users who have also transmitted in the same RB (from both in-cell and out-of-cell users).
SINR can thus be evaluated as in \eqref{eqn_sinr} for all users. 
The SINR can be calculated by plugging in the channel estimates as detailed in Theorem \ref{thm_sinr}.

\section{Proof of Theorem \ref{thm_sinr_mrc_deteq}} \label{appendix_sinr_mrc_deteq}
As the number of antennas gets large, both $\| \hat{\mathbf{h}}_{tqm}^{qk} \|^2$ and $| \hat{\mathbf{h}}_{tqm}^{qkH} \hat{\mathbf{h}}_{tji}^{qk} |^2$ converge almost surely (a.s.) to their deterministic equivalents~\cite{ref_couillet_rmt_2011}. 
Evaluating the deterministic equivalents as in \cite{ref_couillet_rmt_2011} and plugging into the SINR expression in place of the original terms, we can find an approximation to the SINR in the high antenna regime.
As $N$ gets large, the SINR with maximal ratio combining converges almost surely (${\rho}_{tqm}^k \stackrel{\text{a.s.}}{\longrightarrow} \overline{\rho}_{tqm}^k$) to
\begin{align*}
\overline{\rho}_{tqm}^k &= \frac{ {\tt Sig}_{tqm}^k }{\epsilon_{tqm}^k ( N_0 + {\tt IntNC}_{tqm}^k ) + {\tt IntC}_{tqm}^k},
\end{align*}
where ${\tt Sig}_{tqm}^k$ is the desired signal, $ {\tt IntNC}_{tqm}^k$ represents the non-coherent interference,  and ${\tt IntC}_{tqm}^k$ represents the coherent interference.
These can be evaluated as
\begin{align*}
\epsilon_{tqm}^k  \!\! &= \!\! 
\left(\!\!  \begin{array}{l}
N_0 \|\mathbf{c}_{tqm}^{qk}\|^2 
+  \sum\nolimits_{n \in \mathcal{S}_{kj}} |\mathbf{p}_{qn}^H \mathbf{c}_{tqm}^{qk} |^2  g_{tqn} \beta_{qn}^q \sigma_h^2 \\
+  \sum\nolimits_{l \in \mathcal{Q}^q} \sum\nolimits_{n \in \mathcal{S}_{1j}} |\mathbf{p}_{ln}^H \mathbf{c}_{tqm}^{qk} |^2  g_{tln} \beta_{ln}^q \sigma_h^2
\end{array} \!\!  \right) \!\!,  \\
{\tt Sig}_{tqm}^k  \!\! &= \!\! N p_{qm} g_{tqm} (\epsilon_{tqm}^k)^2 \!\!, \\
{\tt IntNC}_{tqm}^k \!\! &=\!\!  \left( \!\! \begin{array}{l}
p_{qm} g_{tqm} \delta_{tqm}^{qk} +  \sum\nolimits_{n \in \mathcal{S}_{kj}}  p_{qn} g_{tqn} \beta_{qn}^q \sigma_h^2 \\
+  \sum\nolimits_{l \in \mathcal{Q}^q} \sum\nolimits_{n \in \mathcal{S}_{1j}} p_{ln} g_{tln} \beta_{ln}^q \sigma_h^2
\end{array} \!\! \right)  \!\!,  \\
{\tt IntC}_{tqm}^k \!\! &= \!\! N  \!
\left( \!\! \! \begin{array}{l} 
\sum\nolimits_{n \in \mathcal{S}_{kj}} |\mathbf{p}_{qn}^H \mathbf{c}_{tqm}^{qk} |^2  p_{qn} g_{tqn} \beta_{qn}^{q2} \sigma_h^4 \\
+  \sum\nolimits_{l \in \mathcal{Q}^q} \sum\nolimits_{n \in \mathcal{S}_{1j}} |\mathbf{p}_{ln}^H \mathbf{c}_{tqm}^{qk} |^2 p_{ln} g_{tln} \beta_{ln}^{q2} \sigma_h^4
\end{array} \!\! \! \right)  \!\!.
\end{align*}
Here,  $\delta_{tqm}^{qk} $ and ${\mathbf{c}}_{tqm}^{qk}$ are obtained from Theorems \ref{thm_ch_est} and~\ref{thm_sinr}, respectively, for the three estimation schemes.
The above expressions are obtained by setting $\mathbf{a}_{tqm}^{k} = \hat{\mathbf{h}}_{tqm}^{qk}$~\cite{ref_bjornson_mimo_2017} and replacing each of the terms involving $\hat{\mathbf{h}}_{tji}^{qk}$ in \eqref{eqn_sinr} with their respective deterministic equivalents.

\end{document}